\documentclass[a4paper,fleqn,usenatbib]{mnras}

\usepackage[T1]{fontenc}
\usepackage{ae,aecompl}

\usepackage{graphicx}	
\usepackage{amsmath}	
\usepackage{amssymb}	

\newcommand{\be}{\begin{equation}}
\newcommand{\ee}{\end{equation}}
\newcommand{\bea}{\begin{eqnarray}}
\newcommand{\eea}{\end{eqnarray}}

\title[AGN in Clusters and Field]{NoSOCS in SDSS. VI. 
The Environmental Dependence of AGN in Clusters and 
Field in the Local Universe}
\author[Lopes, Ribeiro \& Rembold]{P. A. A. Lopes$^{1,2}$\thanks{E-mail: 
    plopes@astro.ufrj.br}, A. L. B. Ribeiro$^{3}$, S. B. Rembold$^{4}$\\
\\
$^{1}$Observat\'orio do Valongo, Universidade Federal do Rio de Janeiro, 
Ladeira do Pedro Ant\^onio 43, Rio de Janeiro, RJ, 20080-090, Brazil\\
$^{2}$ Institut d'Astrophysique de Paris (UMR 7095: CNRS \& UPMC, Sorbonne Universit\'es), 98 bis Bd Arago, F-75014 Paris, France\\ 
$^{3}$Laborat\'orio de Astrof\'isica Te\'orica e Observacional -- Departamento 
de Ci\^encias Exatas e Tecnol\'ogicas -- Universidade Estadual de Santa Cruz,\\
45650-000, Ilh\'eus, BA, Brazil\\
$^{4}$Universidade Federal de Santa Maria -- 97105-900, Santa Maria-RS, Brazil\\
}

\date{Accepted 2017 August 4. Received 2017 August 4; in original form 2017 June 13.}

\pubyear{2017}

\begin{document}
\label{firstpage}
\pagerange{\pageref{firstpage}--\pageref{lastpage}}
\maketitle

\begin{abstract}
  We investigated the variation in the fraction of optical active galactic
  nuclei (AGN) hosts with stellar mass, as well
  as their local and global environments.
  Our sample is composed of cluster members and
  field galaxies at $z \le 0.1$ and we consider only strong AGN.
  We find a strong variation in the AGN fraction
  (F\textsubscript{AGN})
  with stellar mass. The field population comprises
  a higher AGN fraction compared
  to the global cluster population, especially for
  objects with log $M_* > 10.6$. Hence, we restricted our analysis to more
  massive objects. We detected a smooth variation in the
  F\textsubscript{AGN} with local stellar mass density for cluster objects,
  reaching a
  plateau in the field environment. As a function of clustercentric distance
  we verify that F\textsubscript{AGN} is roughly constant for R $> $ R$_{200}$,
  but show a
  steep decline inwards. We have also verified the dependence of the AGN
  population on cluster velocity dispersion,
  finding a constant behavior for low mass systems
  ($\sigma_P \lesssim 650-700$ km s$^{-1}$). However, there is a strong decline
  in F\textsubscript{AGN} for higher mass clusters ($>$ 700 km s$^{-1}$).
  When comparing the F\textsubscript{AGN}
  in clusters with or without substructure we only find different results
  for objects at large radii (R $> $ R$_{200}$), in the sense that clusters
  with substructure present some excess in the AGN fraction.
  Finally, we
  have found that the phase-space distribution of AGN cluster members is
  significantly different than other
  populations. Due to the
  environmental dependence of F\textsubscript{AGN} and their phase-space
  distribution we interpret AGN to be the result of galaxy interactions,
  favored in environments where the relative velocities are low, typical
  of the field, low mass groups or cluster outskirts.
\end{abstract}




\section{Introduction}

A key issue in the understanding of galaxy formation and evolution is
related to the triggering mechanisms of active galactic nuclei. Although
supermassive black holes (SMBHs) are common in the central parts of
virtually all massive galaxies, it is not clear why just
a few objects display strong nuclear
activity due to matter accretion. Different mechanisms are possible to
feed the inflow of gas to the galaxy centers activating their nuclei.
Among those are major and minor mergers, bar influence, disc instability and
tidal effects \citep{moo96, elm98, spr05, gen08}. Some
of those (such as mergers and interactions)
are also known for accelerating the star formation rate (SFR), pointing to
a co-evolution between black hole growth and star formation activity in
galaxies \citep{sil08, hic14, aza15, wan15, alb16}.

As galaxy interactions and mergers are environment related processes the
investigation of the nuclear activity dependence on environment may be
paramount on providing clues on the AGN triggering. Hence, the investigation
of the AGN properties and frequency in the field, groups and clusters can
be important for understanding those system's evolution. Several studies
tackle this issue, finding in general an anti-correlation between AGN
fraction and environmental density, so that fewer
AGN are found in clusters compared
to the field \citep{dre99, kau04, alo07}. Discrepancies among
different studies are generally
explained by the use of different samples (in terms of luminosity and
mass, and wavelength selection), and environment definitions.
Regarding the use of AGN selected in different wavelengths the
environmental dependence and triggering mechanism for AGN selected
in the radio may be different
from optical and mid-IR selected AGN \citep{sat14, ell15}, as well
as X-rays \citep{mar13, gou14}.

Besides the local environment, the global may also be relevant to the
AGN population. For instance, \citet{pop06} find an anti-correlation
between AGN fraction and cluster velocity dispersion, so that 
F\textsubscript{AGN} decreases as $\sigma_P$ increases. The authors
interpret this result as a consequence of the galaxy-galaxy merger
inefficiency in clusters, as cluster galaxies have high relative velocities.
Hence, the AGN phenomenon is more common in the field and low mass groups.
The global environment can also be indicated by how long a galaxy inhabit 
a cluster, as recent arrivals have been less affected by the cluster
potential. We can estimate how long since a galaxy infall from
their location in the cluster phase-space \citep{oma13}. Hence,
we can investigate the AGN distribution in phase-space trying to
understand their evolution within clusters.

Finally, another important factor on assessing the environment influence
(actually the cluster environment) is the dynamical stage of such systems.
\citet{pim13} investigate the AGN fraction in six nearby relaxed
clusters. They use only systems with no substructure signs as they argue
mergers of groups and clusters may locally boost AGN activity, biasing
the analysis. \citet{pop06} also avoid clusters with substructure,
but for a different reason. As they investigate the
F\textsubscript{AGN}-$\sigma_P$ connection they did not want to have
systems with possible wrong values for $\sigma_P$.

Nonetheless, despite what is said above, it is essential
  to emphasize the variation of
  F\textsubscript{AGN} with environment is not a consensus in the literature,
  as there are many studies pointing to no environment connection in the
  number of AGN found. For instance, \citet{mil03} find that the fraction of
  galaxies hosting AGN does not change from the central parts of clusters
  all the way to the field. \citet{von10} results also point to an
  independence of F\textsubscript{AGN} and clustercentric radius, but only for
  powerful AGN hosted by star-forming galaxies. The same is not true for
  weak optical AGN hosted by red galaxies. The lack of environment influence
  is also supported by the work of \citet{sab15}, who find the local
  density and one-to-one interactions to have little impact on the AGN
  activity. \citet{del17} also point to no environmental dependence on
  F\textsubscript{AGN} for objects in the region of the Abell 901/2
  multi-cluster system.

This work is the sixth of a series aiming to investigate cluster and
galaxies' properties at low redshifts ($z \le 0.1$). On this current paper
we focus on the investigation of the
variation of the AGN population as function of environment and stellar mass.
Our data set is composed of objects in two complementary
luminosity ranges ($M_r \le M^*+1$ and $M^*+1 < M_r \le M^*+3$), where $M^*$
is the characteristic magnitude of the luminosity function in the $r-$band.
However, in the current work, as explained in $\S$ \ref{agn-smass},
we apply a stellar mass cut
(log $M_* > 10.6$) so that nearly all objects are in the bright regime.

We organized this paper in the following way: $\S$2 has the data
description, where we define the cluster and field samples, and
discuss the local galaxy density estimates, the
stellar population properties, and the
AGN sample. In $\S$3 we present the relation between AGN fraction and
stellar mass. The environmental variation of the AGN population is
investigated in $\S$4. We search for the connection to the local
and global environment, indicated by cluster velocity dispersion
and dynamical stage. The possible dependence of
the environmental variation
on galaxy host type is searched in $\S$5. In $\S$6 we compared the AGN
distribution in phase-space to other three populations. Our main results are
summarize in $\S$7.  The cosmology assumed in this work considers
$\Omega_{\rm m}=$0.3,  $\Omega_{\lambda}=$0.7, and
H$_0 = 100$ $\rm h$ $\rm km$ $s^{-1}$ Mpc$^{-1}$, with $\rm h$ set to 0.7.
For simplicity, in the following we are going to use the term ``cluster''
to refer loosely to groups and clusters of galaxies.

\section{Data}

This work is based on a sample of 6415 galaxies belonging to 152 groups and
clusters and 4676 selected as field galaxies. Our clusters span the range
$150 \la \sigma_P \la 950$ km s$^{-1}$, or the equivalently in terms
of mass, $10^{13} \la M_{200} \la 10^{15} M_{\odot}$. The cluster and field
samples are restricted to $z \le 0.100$. In the next two subsections we
  briefly describe the cluster and field samples. We also describe the main
  galaxy properties used in this paper, as well as the AGN selection and
  the local density estimates. Further details on the construction
of those two samples can be found in the previous papers of this series
(papers I to V).

\subsection{Cluster Sample} \label{cls-sample}

The cluster sample was originally selected from the digitized version of the
Second Palomar Observatory Sky Survey (POSS-II; DPOSS,
\citealt{djo03, gal04, ode04}), and it is named the Northern Sky 
Optical Cluster Survey (NoSOCS, \citealt{gal03, lop04, gal09}).
In paper I we defined this low-$z$ cluster sample from the supplemental
version of the NoSOCS \citep{lop04}, re-estimating photometric
redshifts as in \citet{lop07}. This
low-redshift sample was complemented with more massive systems
from the Cluster Infall Regions in SDSS (CIRS) sample (\citealt{rin06},
hereafter RD06). Most of the galaxy and cluster properties, including the
local density estimates (for field and cluster galaxies) were derived
using the 7th Sloan Digital Sky Survey (SDSS) release (DR7). The exception is
given by the stellar population properties (see below) that were derived from
the SDSS DR8. 

The redshift limit of the sample ($z = 0.100$) is due to incompleteness in
the SDSS spectroscopic survey for higher redshifts, where galaxies fainter 
than $M^*+1$ are missed, biasing the dynamical analysis (see discussion in
section 4.3 of \citealt{lop09a}). We eliminated interlopers and selected
cluster members using the ``shifting gapper'' technique \citep{fad96, lop09a},
applied to all galaxies with spectra available within a maximum aperture of
2.50 h$^{-1}$ Mpc, and within $\pm$ 4000 km s$^{-1}$ of the cluster
redshift. After selecting cluster members we perform a virial
analysis, obtaining estimates of velocity dispersion, physical radius and
mass ($\sigma_P$, $R_{500}$, $R_{200}$, $M_{500}$ and $M_{200}$; details in
paper I).

For the clusters with at least five galaxy members within R$_{200}$ we also
have a substructure estimate, based on the DS (or $\Delta$) test
\citep{dre88}. Another substructure estimate was derived from
  the Hellinger Distance (HD) measure, but in that case we impose a
  minimum of 20 galaxies within R$_{200}$ (see \citealt{rib13b, dec17}).
We also estimated X-ray luminosity ($L_X$, using ROSAT All Sky Survey
data), optical luminosity ($L_{opt}$) and richness (N$_{gals}$,
\citealt{lop09a, lop09b}). The centroid of each NoSOCS cluster is a
luminosity weighted estimate, which correlates well with the X-ray peak
(see \citealt{lop06}).

\subsection{Field Sample} \label{fld-sample}

  The galaxy field sample is constructed in an independent manner of the cluster
  sample above, meaning it is not taken as the interlopers found within the
  maximum sampling cluster regions (2.50 h$^{-1}$ Mpc, and within
  $\pm$ 4000 km s$^{-1}$). As described in \citet{lop14, lop16} we select as
  field objects those galaxies from the whole SDSS DR7 that are not associated
  to a group or cluster. Our cluster catalog used as reference is the one from
  \citet{gal09}, containing more than 15,000 cluster candidates over
  11,411 deg$^2$. Our approach is very conservative, so that a galaxy is said
  to belong to the \emph{field} if it is not found within 4.0 Mpc and not
  having a  redshift offset smaller than 0.06 of any cluster from
  \citet{gal09}. Note the clusters in that catalog have a photometric
  (not spectroscopic) redshift estimate. We select more than 60,000 field
  galaxies, but work with a smaller subset (randomly chosen). The field
  sample we used has 2,936 galaxies at $z \le$ 0.100 with $M_r \le M^*+1$,
  and 1,740 at $z \le$ 0.045 with $M^*+1 < M_r \le M^*+3$. The local
  density is estimated for those objects (see below) in the same way
  as done for the cluster members.

  It is important to stress this field sample is based on a comparison to
  one cluster catalog \citep{gal09}. Cluster samples are generally complete
  for rich systems, but not for the smaller mass groups and clusters. Due
  to that some field objects may actually belong to small groups, that are
  not listed by \citet{gal09} (and could also be missing from other cluster
  catalogs). The number of ``field objects'' that could be group members is
  expected to be small, and most importantly, their local densities are
  much smaller than typical values of the central regions of groups and
  clusters, being at most comparable to the values found in the outskirts of
  those systems.

\subsection{Absolute Magnitudes and Colours} \label{abs-mag-col}

In the current work we consider the stacked properties of galaxies
  in our sample. That is done using the radial offset in units of $R_{200}$
  (only for member galaxies), absolute magnitudes, colours and local densities
  of all galaxies in our cluster and field samples. The absolute magnitudes
  of each galaxy in five SDSS bands ($ugriz$) are derived using the
  formula: $M_x = m_x - DM - kcorr - Qz$ ($x$ is one of the five SDSS
  bands we used), DM is the distance modulus (using the galaxy redshift),
  $kcorr$ is the k$-$correction and $Qz$ ($Q = -1.4$, \citealt {yee99}) is
  a mild evolutionary correction applied to the magnitudes. Rest-frame
  colours are also derived for all objects. The magnitudes
  we obtained from the SDSS are de-reddened model magnitudes (see paper I).
  We consider the k$-$corrections available in the SDSS database, for each
  galaxy and band.

\subsection{The stellar population properties}

A variety of galaxy properties were obtained by different research groups
  for the SDSS. These parameters consider galaxy spectra or the broad band
  galaxy photometry, and are derived from spectral energy distribution (SED)
  fitting of stellar population synthesis models. In the current work we
  consider parameters derived from the ``galSpec'' analysis provided by
  the MPA-JHU group (from the Max Planck Institute for Astrophysics and
  the Johns Hopkins University; \citealt{bri04}). A brief description of
  those parameters is given below.

\subsubsection{MPA-JHU} \label{mpa-jhu}

The galaxy properties from MPA-JHU, named ``galSpec'', were obtained for
  DR8 galaxy spectra (nearly all of which were in DR7). The galaxy parameters
  we used in the current work are the BPT classification \citep{bpt81}, the
  stellar mass and the star formation rate. From the BPT diagram the MPA-JHU
  group classified galaxies into the following categories: ``Star Forming'',
  ``Composite'', ``AGN'', ``Low S/N Star Forming'', ``Low S/N AGN'', and
  ``Unclassifiable''.
This classification using the BPT diagram considers four emission lines
  (see Fig. 1 of \citealt{bri04}) and requires S/N $> 3$ for all lines. In
  particular a ``Low S/N AGN'' has
  [N\textsubscript{II}]6584/H${\alpha}$ $> 0.6$
  (and S/N $> 3$ in both lines), but has [O\textsubscript{III}]5007
  and/or H${\beta}$ with low S/N. ``Low S/N Star Forming'' galaxies are
  those with S/N $> 2$ in H${\alpha}$ after most galaxies with an AGN
  contribution to their spectra are removed.

Besides the galaxy classification described above, we also consider
  the total stellar mass and star formation rate values (SFRs) from
  ``galSpec''. The stellar masses are based on model magnitudes.
  The SFRs are computed within the galaxy fiber aperture and are based
  on the nebular emission lines as described in \citet{bri04}. The galaxy
  photometry is used outside of the fiber \citep{sal07}. For AGN and
  galaxies with weak emission lines the SFRs estimates are derived from
  the photometry. Due to the minimum criteria required by the MPA-JHU
  group ({\it {e.g.}}, redshift, S/N) not all galaxies in our sample are
  matched to the MPA-JHU data set. Hence, our sample with MPA-JHU values
  have 4,953 bright and 1,297 faint member galaxies, and also 2,864 bright
  and 1,672 faint field galaxies. That is 97$\%$ of the original sample.

\subsection{The AGN population} \label{agn_pop_mpa-jhu}

As mentioned above the object classification we consider is based on a BPT
diagram. In the current study we investigate the dependence of the AGN
population on stellar mass, local and global environment. The latter is
traced by the parent cluster mass and can be also studied by comparing
field and cluster objects. We note
the AGN class from ``galSpec'' already excludes LINERs, objects that are not
primary powered by nuclear activity \citep{sch10, sin13}. Objects
  named ``Low S/N Star Forming'' represent a different class (as described
  above) and are not part of our sample. To
further reduce contamination from objects not mainly dominated by nuclear
activity we do not include ``Composite'' objects in our AGN sample. Hence,
our list of AGN considers only objects classified as ``AGN'' (excluding LINERs)
from ``galSpec'' (see \S \ref{mpa-jhu} above). Note that some authors
also include ``Composite'' objects \citep{soh13, bit15} leading to higher
AGN fractions than what we show below.

\subsection {Local Galaxy Density Estimates}

For the current work we adopt the $\Sigma_5$ local galaxy density
  estimator. That is motivated by the conclusions of \citet{mul12} who
  find that nearest neighbor methods are superior on tracing the local
  environment. The choice of the rank of the density-defining neighbor ($n$)
  is also important. We chose to work with $n = 5$ since that is typically
  smaller than the number of galaxies we have per cluster and is a common
  estimate in the literature. The local galaxy density estimates are derived
  as follows. For each galaxy in our sample we compute the projected
  distance, d$_5$, to the 5th nearest galaxy around it. We also impose to
  the neighbor search a maximum velocity offset of 1000 $km$ $s^{-1}$, and
  a maximum luminosity, which we adopt as $M^* + 1.0$. The local density
  $\Sigma_5$  is simply given by 5/$\pi$d$_5^{2}$, and is measured in  units
  of galaxies/Mpc$^2$. Finally, we also take in account the fiber collision
  issue when deriving galaxy densities. The procedure is well described in
  \citet{lab10, lop14}.

\subsubsection{Stellar Mass Density Estimates}

\begin{figure}
\begin{center}
\leavevmode
\includegraphics[width=3.5in]{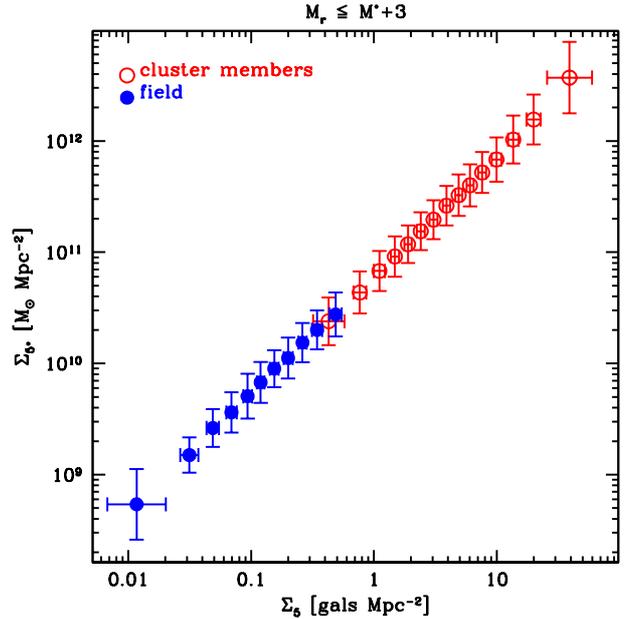}
\end{center}
\caption{Comparison between the galaxy stellar mass density ($\Sigma_{5*}$) and
  the galaxy number density ($\Sigma_5$), for cluster (red open circles) and
  field (blue filled symbols) galaxies. The error bars indicate the 1$\sigma$
  standard error on the biweight location estimate.}
\label{fig:stellarmass_number_density}
\end{figure}

It is well know that the central parts of clusters - normally the most dense
regions in the Universe - are inhabited by the most luminous and massive
galaxies, so that a correlation between local galaxy number density and
stellar mass is expected. Nonetheless, the scatter on this relation may
not be negligible and the environmental dependence of galaxy properties
may be better traced by the the galaxy stellar mass density, instead of
simply the galaxy number density. Hence, we also computed the former. We
adopt the same procedure as above, but added the stellar mass of the
five nearest neighbors, instead of their numbers. The local stellar mass
density $\Sigma_{5*}$ is then given by S$_5^*$/$\pi$d$_5^{2}$, where S$_5^*$
is the sum of the stellar masses of the five nearest neighbours.
$\Sigma_{5*}$ is measured in  units of M$_{\odot}$/Mpc$^2$. In
Fig.~\ref{fig:stellarmass_number_density} we display a comparison
of the galaxy stellar mass density ($\Sigma_{5*}$) {\it vs} the numerical
density ($\Sigma_5$) for the cluster and field galaxies. We can see a good
correlation between the two estimates, stretching from the low density field
environment to the high densities of cluster cores. Hence, for the
current work we adopted the local galaxy stellar mass density ($\Sigma_{5*}$)
as the tracer of the local environment. The error bars in the figure
  indicate the 1$\sigma$ standard error on the ``biweight location estimate'',
  which is a resistant and robust estimator. It is known to be superior to
  the mean and median estimates, both in terms of resistance and robustness.
  See \citet{bee90} for a proper definition of the biweight estimator, as
  well for a comparison of its performance to other estimators.

\section{The AGN Fraction by Stellar Mass} \label{agn-smass}

\begin{figure}
\begin{center}
\leavevmode
\includegraphics[width=3.5in]{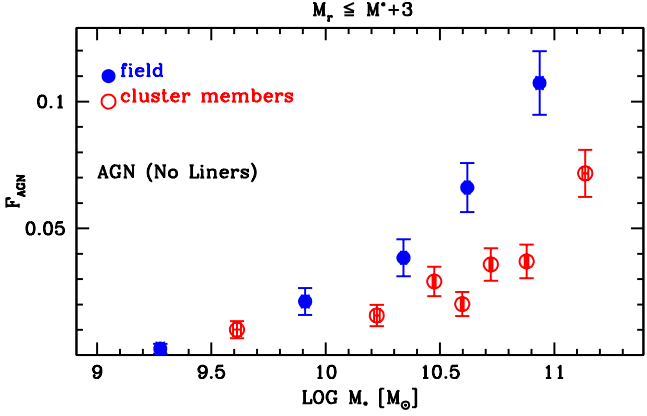}
\end{center}
\caption{Variation of AGN fraction as a function of galaxy stellar mass, for
  cluster (red open circles) and field (blue filled circles) objects. The
  error bars indicate the 1$\sigma$ standard error on the biweight
  location estimate.}
\label{fig:agn_mstar_bins}
\end{figure}

We exhibit in Fig.~\ref{fig:agn_mstar_bins} the variation of the AGN fraction
(F\textsubscript{AGN}) of our sample as a function of galaxy stellar mass,
for cluster and field objects. F\textsubscript{AGN} is defined here as the
number of objects classified as AGN ($\S$\ref{agn_pop_mpa-jhu}) divided
by the total number of galaxies in our sample (no luminosity or morphological
cuts are applied). As expected there is a clear increase in the
number of AGN as we
go from low to high-mass objects \citep{bes05, bru09, pim13}.
However, it is interesting to note that the AGN
fraction is normally higher in the field than in clusters, especially
for log $M_* \gtrsim 10.6$. Hence, from this figure we can also detect
the impact of global environment, as at fixed stellar mass the
fraction of AGN decreases as we move from the field to clusters.
Due to possible incompleteness in the
AGN selection for low luminosity (massive) objects, we decided to consider
only massive objects (log $M_* > 10.6$) in the current work. Taking this
limit we also enforce a clearer separation between field and clusters.
Another important reason to drop objects with log $M_* \le 10.6$ is the fact
that there is no strong environmental variation in the AGN fraction for those
low mass objects \citep{pim13}. We have also found that. Hence, we
chose to exclude the low mass regime as its inclusion weakens the
connection between AGN fraction and environment (see below).

After applying the stellar mass cut the
sample decreases to 3,118 member galaxies and 1,561
field galaxies. That represents $\sim$ 50\% of the cluster sample and
$\sim$ 34\% of the field sample described in the end of $\S$ \ref{mpa-jhu}.
However, this stellar mass cut effectively is a luminosity cut, so that the
vast majority of objects ($> $ 99\%) surviving is in the bright regime
($M_r \le M^*+1$; $\S$ \ref{abs-mag-col}). Considering
the bright samples described in $\S$ \ref{mpa-jhu} (4,953 members and 1,672
field objects), after applying the stellar mass cut we ended up with
$\sim$ 63\% of those bright cluster members and $\sim$ 54\% of the bright
field objects. Regarding the global AGN fractions, in the cluster and
  field environment, after we apply the stellar mass cut we have the
  following: out of the 3,118 massive cluster members we have 133
  ($\sim$ 4,3\%) which are AGN. In the field, out of the 1,561 massive
  objects we find 139 ($\sim$ 8,9\%) AGN.

\section{The Environmental Variation of the AGN Population}

\subsection{Local Environment Dependence}

\subsubsection{Variation with Galaxy Stellar Mass Density}

\begin{figure}
\begin{center}
\leavevmode
\includegraphics[width=3.5in]{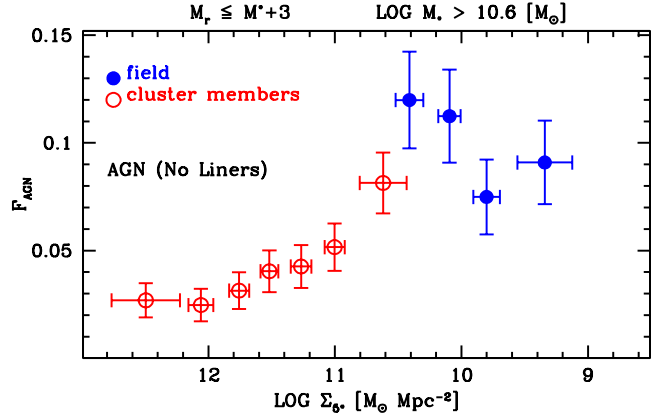}
\end{center}
\caption{Dependence of AGN fraction as a function of local galaxy stellar
  mass density ($\Sigma_{5*}$), for cluster (red open circles) and 
  field (blue filled circles) objects. The error bars indicate the 1$\sigma$
  standard error on the biweight location estimate.}
\label{fig:agn_msden_bins}
\end{figure}

In this section, we investigate the variation of the AGN fraction with
local environment, given by the galaxy stellar mass density. In
Fig.~\ref{fig:agn_msden_bins} we show the variation in the relative number
of AGN as a function of local galaxy stellar mass density ($\Sigma_{5*}$).
From the low to the high density regime in the field we detect a small
increase in the fraction of AGN. However, from the outskirts of clusters
(low densities) to their cores (high density environment) we can see
a strong decrease in the AGN fraction as a function of local stellar mass
density. A possible explanation for those results is that nuclear activity
is triggered by mergers, which may be preferable in the higher density field
environment or within low mass groups and the outskirts of clusters, which
is the case for 10.0 $< $ log $\Sigma_{5*} <$ 10.6. In such low $\Sigma_{5*}$
regions the relative galaxy velocities are small, favouring galaxy mergers
(see $\S$4.2.1 below).

\subsubsection{Variation with Clustercentric Distance}

\begin{figure}
\begin{center}
\leavevmode
\includegraphics[width=3.5in]{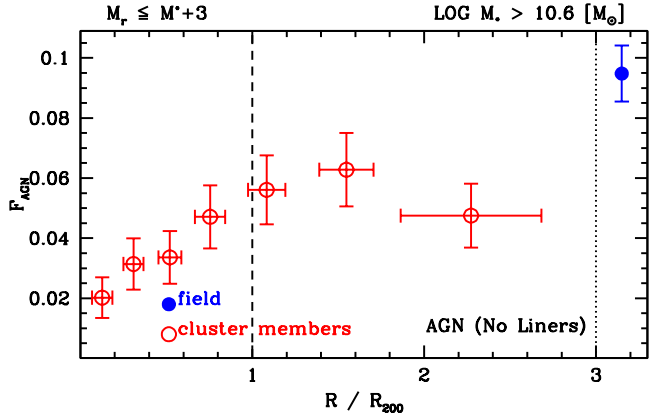}
\end{center}
\caption{Variation of AGN fraction as a function of clustercentric distance
  for the cluster objects (red open circles). The corresponding 
  estimate for the field is indicated by the blue filled circle. The
  error bars indicate the 1$\sigma$ standard error on the biweight location
  estimate.}
\label{fig:agn_rad_bins}
\end{figure}

In Fig.~\ref{fig:agn_rad_bins} we display the relation between the AGN
fraction and clustercentric distance for cluster members. The field value
is also displayed as a reference. We can see a clear variation in
F\textsubscript{AGN} as a function of normalized radial offset. Inside
R$_{200}$ there is a steep decline in the AGN fraction. However, outside
R$_{200}$ the AGN fraction is nearly constant. Two more important
results can still be seen on this figure. First we see the field value is
significantly larger than the fractions within clusters, even in their
outskirts. Second, there is a small decrease in the AGN fraction for
R $> $ 1.5R$_{200}$, but that is not significant, and the fraction at
R $> $ 2R$_{200}$ is at least two times larger than found in the cluster
cores. \citet{pim13} mention it is common to find in the literature
results pointing to no difference between cluster and field AGN fractions
(but see the results of \citealt{soh13} for different conclusions).
\citet{pim13} also claim no significant difference between the
results in the core and at R $\sim $ 2R$_{200}$. One possible explanation
for those discrepancies is the mix of galaxy host types. \citet{soh13}
find a significant difference between field and clusters, but mainly for
AGN hosted by ETGs. That is corroborated by
Fig.~\ref{fig:agn_rad-bins-etg-ltg} ($\S$5). The other reason,
possibly the most important, is the inclusion of low mass objects. We
notice that low mas objects (log $M_* \le$ 10.6) display similar AGN
fractions in clusters and the field. The difference in the central and
outskirts of clusters is also not significant. That can also be seen in
Fig.~6 of \citet{pim13}.

\subsection{Global Environment Dependence}

\subsubsection{Connection to Cluster Velocity Dispersion}

\begin{figure}
\begin{center}
\leavevmode
\includegraphics[width=3.5in]{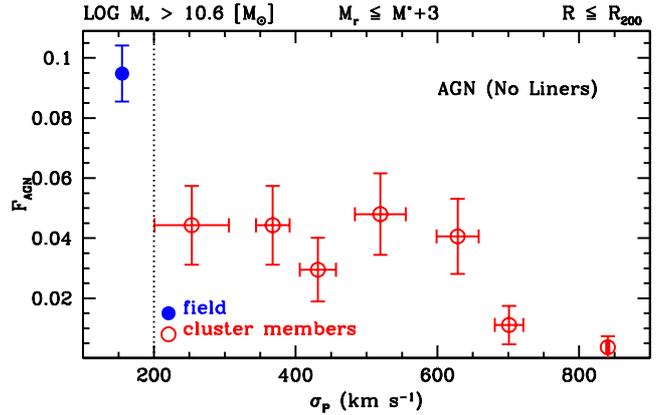}
\end{center}
\caption{Relation between the AGN fraction and the velocity dispersion of the
  parent cluster (red open circles). The corresponding 
  estimate for the field is indicated by the blue filled circle. The
  error bars indicate the 1$\sigma$ standard error on the biweight location
  estimate.}
\label{fig:agn_vdisp_bins}
\end{figure}

\begin{figure}
\begin{center}
\leavevmode
\includegraphics[width=3.5in]{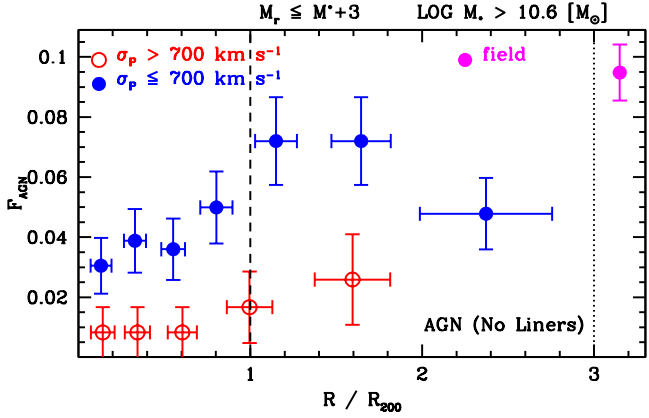}
\end{center}
\caption{Variation of AGN fraction as a function of clustercentric distance
  for low ($\sigma_P \le 700$ km s$^{-1}$) mass clusters (displayed by blue
  filled circles) and high mass systems ($\sigma_P > 700$ km s$^{-1}$, red
  open circles). The corresponding 
  estimate for the field is indicated by the magenta filled circle. The
  error bars indicate the 1$\sigma$ standard error on the biweight location
  estimate.}
\label{fig:agn_rad-vdisp_bins}
\end{figure}

We examine the dependence of F\textsubscript{AGN} on the parent cluster
mass, indicated by the velocity dispersion. That is shown in
Fig.~\ref{fig:agn_vdisp_bins}. There is a constant relation between
AGN fraction and $\sigma_P$ for low velocity dispersion systems
($\sigma_P \lesssim 650-700$ km s$^{-1}$). For higher mass systems
($>$ 700 km s$^{-1}$) there is a steep decline in F\textsubscript{AGN}.
Hence, the environmental variation is also linked to a dependence on the
parent cluster mass. In our sample of 152 groups and clusters we
  have 12 massive systems ($\sigma_P >$ 700 km s$^{-1}$) and 140 lower mass
  objects ($\sigma_P \le 700$ km s$^{-1}$). The number of massive
  galaxies (log $M_* > 10.6$) in the 12 massive clusters is 596
  ($\sim$ 50 per cluster), being 2,522 for the 140 ($\sim$ 18 per cluster)
  low velocity dispersion systems. We show in
Fig.~\ref{fig:agn_rad-vdisp_bins} the
clustercentric variation of the AGN fraction, but for low and high mass
systems ($\sigma_P \le,> 700$ km s$^{-1}$). At all radii the AGN fractions
in low mass clusters is significantly higher than for massive systems.
Nonetheless, the environmental variation is still present, as in both cases
there is a strong decline once within R$_{200}$.

At first, the lack of F\textsubscript{AGN} variation for the whole halo mass
range could be interpreted as
in contradiction to the F\textsubscript{AGN}$-\sigma_P$ anti-correlation
found by \citet{pop06}. They detect a decrease in F\textsubscript{AGN}
from $\sigma_P \sim 200$ to $\sigma_P \sim 600$ km s$^{-1}$.
F\textsubscript{AGN} is then nearly constant for higher $\sigma_P$ values.
On the contrary, we find a constant relation in the group regime (up to
$\sigma_P \sim 650$ km s$^{-1}$), and an abrupt decrease for higher cluster
masses. Those inconsistent results can be explained by the different samples,
as we only use high mass objects (log $M_* > 10.6$), while \citet{pop06}
includes low luminosity AGN (hosted by lower mass galaxies). It
is also not clear if they consider composite and LINERs among their AGN.
In any case, we verified that when including lower mass objects
(log $M_* \le 10.6$) we also detect a variation in F\textsubscript{AGN}
starting at $\sigma_P \sim 200$ km s$^{-1}$, although it is dominated by the
first bin (composed by the poorest clusters).

\begin{figure}
\begin{center}
\leavevmode
\includegraphics[width=3.5in]{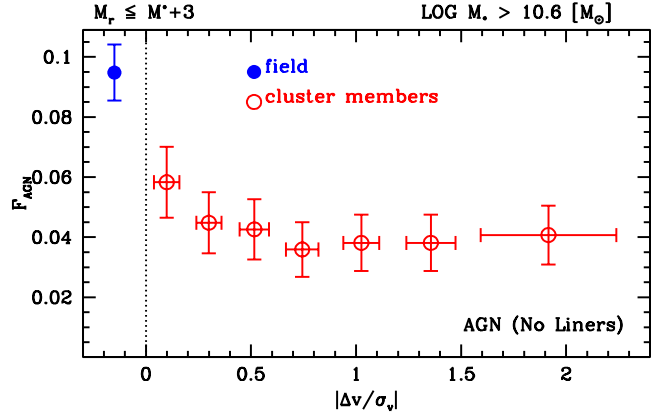}
\end{center}
\caption{Variation of AGN fraction as a function of normalized relative
  velocities ($\Delta_v/\sigma_v$) displayed as red open circles. The
  corresponding estimate for the field is indicated by the blue filled
  circle. The error bars indicate the 1$\sigma$ standard error on the
  biweight location estimate.}
\label{fig:agn_offv_bins}
\end{figure}

As discussed by \citet{pop06} the anti-correlation between AGN
fraction and cluster velocity dispersion may be related to the galaxy-galaxy
merger inefficiency in clusters. The AGN phenomenon would be favoured
in the field and low mass groups or cluster outskirts, as central cluster
galaxies exhibit higher relative velocities than group galaxies or small
field associations. The results from $\S$4.1.1, regarding the local
environment variation, indicate the same conclusions. However, in order
to reinforce that result we show in Fig.~\ref{fig:agn_offv_bins} the
variation of the AGN fraction as a function of the normalized galaxy
velocity offsets.
The offsets are relative to their parent group/cluster redshift and the
normalization considers the velocity dispersion of their parent system. We
can see a smooth decline from the field to clusters, as galaxy velocities
increase, confirming larger AGN fractions for smaller velocities.
Another possible explanation for the anti-correlation between
F\textsubscript{AGN} and $\sigma_P$ is the environmental variation in
AGN galaxy hosts, as investigated in papers III, IV and V. We tackle this
issue in $\S$\ref{agn_morph}.

\subsubsection{Role of the Cluster Dynamical Stage}

We have also investigated if the dynamical stage of the clusters plays a
role in our findings (paper III).
Regarding substructure we classify the clusters using
the $\Delta$ (or DS) test \citep{dre88}
and the Hellinger Distance (HD) measure, used to detect deviations
from a Gaussian velocity distribution \citep{rib13b}. In previous works
\citep{kra13, rib13a} we considered different tests (such as
Anderson-Darling and Kolmogorov-Smirnov)
of gaussianity applied to the galaxy velocity distribution. For the current
work we decided to adopt the HD measure as it is the least vulnerable method to
type I and II statistical errors \citep{rib13b}. We decided to
perform the substructure analysis also with the HD measure as the AGN fraction
may be boosted when $\sigma_P$ is overestimated. In this case, a test based
on galaxy velocities may be even more relevant than a 3D test (such as
the DS).

\begin{figure}
\begin{center}
\leavevmode
\includegraphics[width=3.5in]{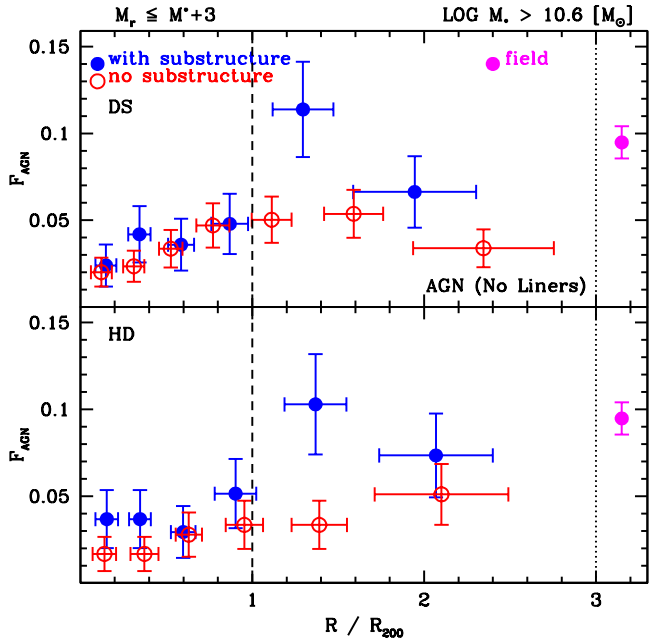}
\end{center}
\caption{Relation between AGN fraction and clustercentric distance
  for systems with no signs of substructure (red open circles) and
  for which a test indicates significant substructure (blue filled circles).
  The corresponding estimate for the field is indicated by the magenta filled
  circle. In the top panel we display the results considering the DS
  substructure test, while in the bottom panel we exhibit the results of
  the HD measure. The error bars indicate the 1$\sigma$ standard error on the
  biweight location estimate.}
\label{fig:agn_rad-sub_bins}
\end{figure}

In Fig.~\ref{fig:agn_rad-sub_bins}
we display the AGN fraction {\it vs} clustercentric distance for systems
with or without substructure. In the top and bottom panels we display the
results obtained with the DS and HD methods, respectively. We can see that
within the virial radius
(approximated by R$_{200}$) there is no difference in the results. A larger
value for F\textsubscript{AGN} for clusters with substructure can only
be seen for R $>$ R$_{200}$ (especially at R $\sim$ 1.5$\times$R$_{200}$).
Hence, we assume no major impact for the inclusion
of systems with substructure. If the merger of subunits increases the
nuclear activity in clusters the effect may be fast enough not to be detected
for objects inhabiting the clusters for longer periods (those within
R$_{200}$). It would be possible to detect an increase in the AGN fraction
for non-relaxed clusters only for objects recent infalling, located
in the clusters' outskirts.

\section{The AGN environmental variation by galaxy type} \label{agn_morph}

It is well know the fraction of galaxies hosting an AGN strongly depends on
galaxy morphology, as well as other parameters, such as colour and stellar
mass \citep{cho09, soh13}. Those effects may impact
the environmental variation of AGN hosted by different types of galaxies.
For instance, \citet{soh13} find an environmental variation of the
AGN fraction only when considering galactic nuclei hosted by early-types.
However, it is important to emphasize their work focus mainly on the AGN
population in compact groups.

In order to check the impact on galaxy morphology on our results, we make
a rough separation between early and late-type galaxies, splitting galaxies
according only to the concentration index \citep{str01, kau04, lop14}.
The concentration index $C$ is defined as the ratio of the radii
  enclosing 90 per cent and 50 per cent of the galaxy light in the r-band,
  $R_{90}/R_{50}$.
We simply call early-type objects (ETGs) those with
C $\ge$ 2.6 and the late-type objects (LTGs) with C $<$ 2.6. Note
  the value of $C$ chosen to split the two populations is in good agreement
  to the literature \citep{kau04}. Different values of $C$ could increase
  the completeness of one population, but at the cost of a higher
  contamination. $C =$ 2.6 represents a good compromise to split early and
  late type objects (see discussion in \citealt{lop14}). There could also be
  the case of transitional galaxies (such as ``red discs'' and
  ``blue spheroids''), but those are rare, as investigated in \citet{lop16}.
  As our AGN sample for massive objects is small we chose not to use a more
detailed separation (using several morphological types, or even more
concentration index bins).

\begin{figure}
\begin{center}
\leavevmode
\includegraphics[width=3.5in]{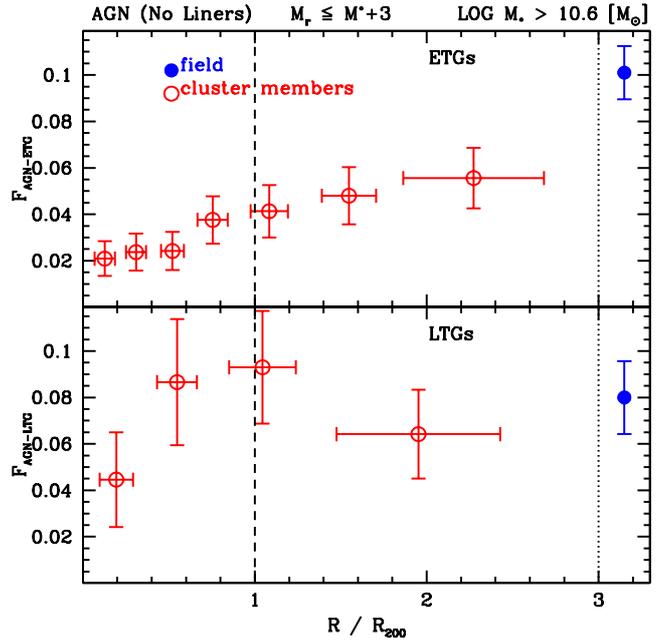}
\end{center}
\caption{Relation between AGN fraction and clustercentric distance
  for AGN hosted by ETGs (top panel) and by LTGs (bottom panel). Those
  results are displayed by red open circles. The corresponding 
  estimate for the field is indicated by the magenta filled circle.
  Those fractions are not relative to all galaxies. Each fraction is
  obtained at fixed morphology, showing the number of AGN hosted by an
  ETG (LTG) divided by the whole ETG (LTG) population. The error bars
  indicate the 1$\sigma$ standard error on the biweight location estimate.}
\label{fig:agn_rad-bins-etg-ltg}
\end{figure}

Fig.~\ref{fig:agn_rad-bins-etg-ltg} display
the relations derived for the AGN fractions for the ETG (top panel) and
LTG (bottom panel) populations. Note that each fraction is not relative to
all galaxies, as in the previous plots. Now we computed the fractions at
fixed morphology, showing the number of AGN hosted by an ETG (LTG) divided
by the whole ETG (LTG) population. The field value is also show as a
reference. We see the environmental variation is present in both cases,
although there is a large scatter for the LTG population. Hence, the AGN
environmental variation we see is not an effect of the morphology-density
relation. Another interesting feature in Fig.~\ref{fig:agn_rad-bins-etg-ltg}
is the fact the AGN fraction in the field for ETGs is larger than the cluster
results, even in their outskirts. That is not the case for the LTG population,
which displays a constant value from the field to R$_{200}$, decreasing
inwards.

As ETGs inhabit clusters for longer periods compared to LTGs its is expected
the fraction of AGN for the first population would be smaller than for the
second. That is the case, if we compare the AGN fractions within R$_{200}$
for the upper and lower panels of Fig.~\ref{fig:agn_rad-bins-etg-ltg}. The
value F\textsubscript{AGN} hosted by LTGs is always larger than for ETGs.

\section{The Location of AGN in phase-space}

\begin{figure*}
\begin{center}
\leavevmode
\includegraphics[width=6.5in]{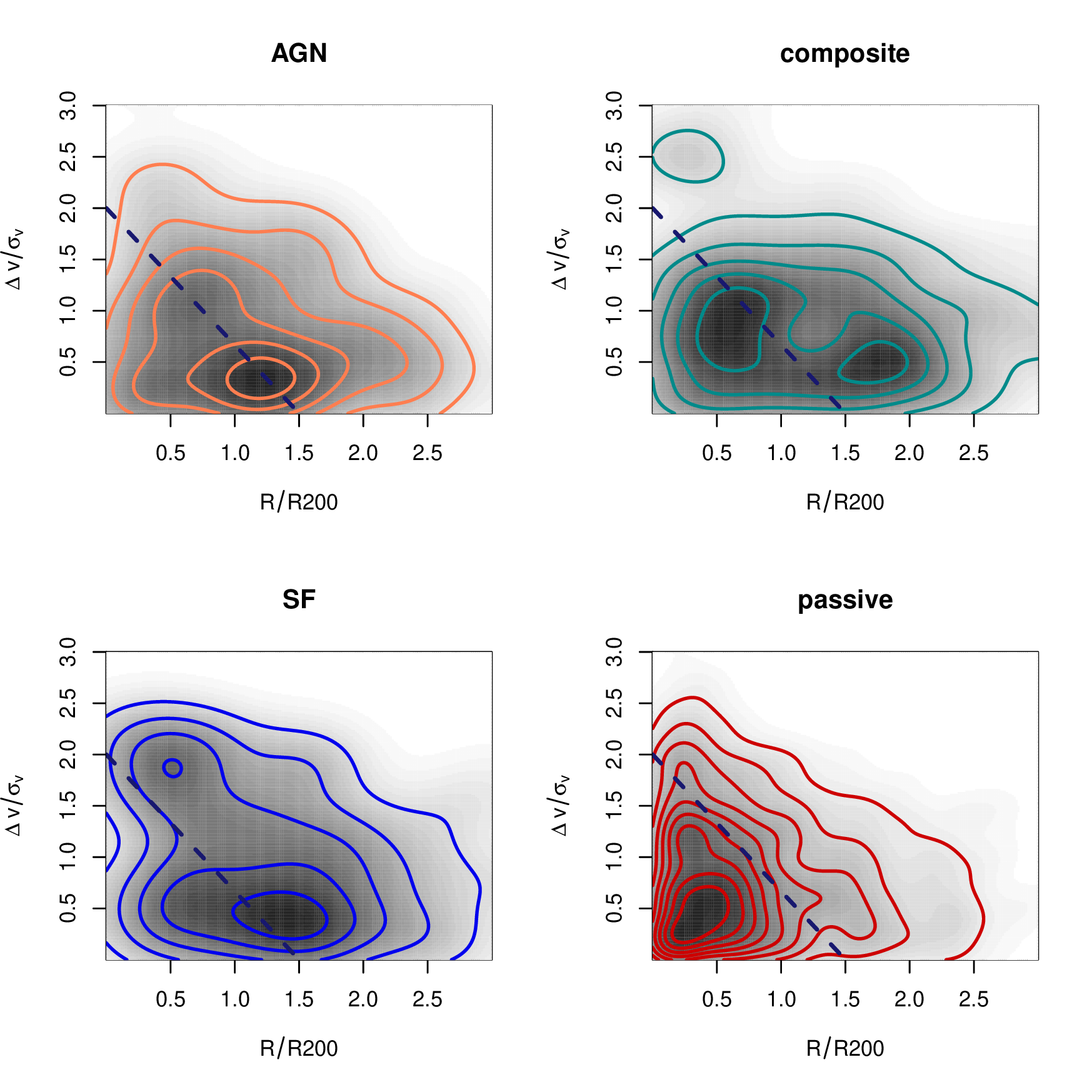}
\end{center}
\caption{The phase-space distribution of AGN (top left), composite
  (top right), SF (bottom left), and passive (bottom right) populations.
  As above, only massive objects (log $M_* > 10.6$) are considered.
  The blue dashed line on each panel represents the equation proposed by
  \citet{oma13} to estimate the infall time of galaxy members (see
  text). This line can roughly be used to separate recent and early infalls.}
\label{fig:four_pops-phase_space}
\end{figure*}

Bearing in mind the last result discussed above it is interesting to
investigate the AGN distribution in phase-space (PS) and compare it to
other galaxy populations. That is also important to check if an enhanced
value of the AGN fractions in the cluster outskirts could be explained
by galaxy interactions (\citealt{pop06, pim13}).
Instead of splitting galaxies according to morphology (using concentration)
we considered four types of objects according to their spectral features.
As mentioned in sections 2.2 and 2.3 we adopted the spectral classification
from the MPA-JHU group. In Fig.~\ref{fig:four_pops-phase_space} we display
the PS distribution of four classes, ``AGN'', ``Composite'',
``Star Forming'' and ``Passive'' (only massive objects,
  log $M_* > 10.6$, are considered). The latter actually is composed of objects
that are ``Unclassifiable''. We loosely  call those objects as passive,
  but that does not come from the spectral classification. However,
  there is good indication those are indeed passive,
  from their location in the SFR$-$M$_*$ plane, as well as in the
  colour-colour diagram [(u-r)$_0$ $\times$ (r-z)$_0$]. The "Unclassifiable"
  population is consistent to passive objects, with $\gtrsim$ 99\% of
  objects falling into the region occupied by passive galaxies on this
  colour-colour plot.

Visually, we infer the AGN population has
different distribution than the composite, SF and passive objects. We
confirm that by running a kernel density based global two-sample comparison
test using the \text{kde} routine of the \text{ks} library in \text{R}.
The \text{kde} test applies the so-called black-box comparisons of
multivariate data \citep{duo12}. The algorithm transforms
data points into kernels and develop a multivariate two-sample test that is
nonparametric and asymptotically normal to directly and quantitatively
compare different distributions. The asymptotic normality bypasses the
computationally intensive calculations used by the usual resampling
techniques to compute the $p$-value. The complete automatic testing procedure
is programmed in the \text{ks} library in \text{R} \citep{duo07}.
In all two sample comparisons we perform the $p$-values of the \text{kde}
test were smaller than 0.05.

On each panel the blue dashed line represents the equation proposed by
\citet{oma13} to estimate the infall time of galaxy members
($\mid\frac{\Delta_v}{\sigma_v}\mid = -\frac{4}{3}\frac{R}{R_{vir}} + 2$).
As pointed out by \citet{agu17} this line can be used to roughly
discriminate recent ($\tau <$ 1 Gyr) and early ($\tau >$ 1 Gyr) infalls.
It is clear from the figure that passive
objects are a majority of early-infalls,
while the SF galaxies are the opposite. The AGN and especially the composite
population are a mix of early and recent infalls. At first, these results
  may seem hard to reconcile to the findings of \citet{pim13}, who
claim no significant different between AGN and other cluster members in the
phase-space (even if restricting to higher mass galaxies). However, it is
worthy remember their sample is restricted to only six clusters, with no
sign of substructure, while we allow the inclusion of disturbed systems.
Most importantly, their sample is restricted only to massive objects, as the
six clusters have $\sigma_v \gtrsim 800$ km s$^{-1}$. As we saw from
Figs.~\ref{fig:agn_vdisp_bins} and \ref{fig:agn_rad-vdisp_bins} the high
mass systems ($\sigma_v > 700$ km s$^{-1}$) show much smaller AGN fractions
than the lower mass objects. The variation with clustercentric distance is
also much less pronounced for the massive clusters.

Considering that AGN are preferentially found outside the virial radius
(R$_{200} <$ R $<$ 1.5$\times$R$_{200}$) and display low normalized relative
velocities ($\mid\Delta_v/\sigma_v\mid$) we interpret those objects may
be triggered by interactions in the outskirts of groups and clusters.
The AGN feedback would then be crucial for quenching SF in members
of groups and clusters \citep{pop06}. As the velocity dispersion
of the galaxy systems grows the AGN production would decrease, leading to the
scenario we find today. That interpretation can also match the discussion
  of \citet{pim13}, who suggest that if AGN are triggered by encounters there
  would be enough time for them to move away from the encounter site. Hence,
  no interaction signs would be detected.

\section{Summary}

In this work we investigated the F\textsubscript{AGN} variation with stellar
mass and environment (both local and global), in the local Universe
($z \le 0.1$). We consider cluster members, as well as a field sample in
the same redshift interval. We also consider only strong AGN (no LINERs in
the sample). After confirming the well know increase in the number of AGN
with stellar mass we restricted our sample to the most massive objects
(log $M_* > 10.6$). The local environment is traced by the local
galaxy stellar mass density and clustercentric distance, while the global
environment is indicated by the cluster potential (traced by its velocity
dispersion). We also searched for any dependence of the fraction of AGN on
the cluster dynamical stage, and also according to galaxy type. Finally,
we compare the locations of different galaxy populations in the cluster
phase-space. Our main results are as follows.

\begin{enumerate}

\item A first indication of the importance of the global environment comes
  from the fact the AGN fraction, at fixed stellar mass, is typically higher
  in the field when compared to clusters
  (Fig.~\ref{fig:agn_mstar_bins}).

\item Regarding the local environment, we detect a roughly constant relation
  between F\textsubscript{AGN} and local galaxy stellar mass density for field
  objects. However, from cluster outskirts inwards we see a steep decline in
  F\textsubscript{AGN} as a function of local stellar mass density
  (Fig.~\ref{fig:agn_msden_bins}).
  
\item A clear dependence between the AGN fraction and clustercentric distance
  is also detected, especially within R$_{200}$ where the variation
  is much stronger (Fig.~\ref{fig:agn_rad_bins}). The field value is also
  larger than the cluster results, even in their outskirts.

\item We find the global environment also plays an important role as we
  detect a dependence on the parent cluster mass, indicated by velocity
  dispersion (Figs.~\ref{fig:agn_vdisp_bins} and
  \ref{fig:agn_rad-vdisp_bins}). We find that F\textsubscript{AGN} is nearly
  constant for $\sigma_P \lesssim 650-700$ km s$^{-1}$, but decreases
  strongly for $>$ 700 km s$^{-1}$.

\item Using two substructure tests we compared the environmental variation
  of clusters with and without substructure (Fig.~\ref{fig:agn_rad-sub_bins}).
  We find no significant difference within R$_{200}$ for these two cases. A
  larger value of the AGN fraction is only present for clusters with
  substructure at R $\sim$ 1.5$\times$R$_{200}$. Hence, we assume there is
  no major impact for using systems with substructure.

\item Using concentration to perform a rough separation between galaxy types
  we find the environmental variation is still detected for both types,
  not being an effect of the morphology-density relation
  (Fig.~\ref{fig:agn_rad-bins-etg-ltg}). However, the field AGN fraction
  (at fixed morphology) is larger than the cluster one, only for ETGs,
  being comparable to the cluster LTGs AGN fraction in the clusters
  outskirts.
  
\item When comparing AGN to other three populations in the cluster
  phase-space (Fig.~\ref{fig:four_pops-phase_space}) we find the AGN
  distribution to be significantly different than the others. 
  
\item The AGN population is a mix of recent and early infalls, being found
  preferentially at R$_{200} <$ R $<$ 1.5$\times$R$_{200}$ with low normalized
  relative velocities ($\mid\Delta_v/\sigma_v\mid$). We interpret the AGN
  phenomenon would be the result of galaxy interactions, which are favoured
  in the field and low mass groups or cluster outskirts, where relative
  velocities are typically low (see also Fig.~\ref{fig:agn_offv_bins}).

\end{enumerate}

\section*{Acknowledgements}

ALBR thanks for the support of CNPq, grants 306870/2010-
0 and 478753/2010-1. PAAL thanks the support of CNPq, grant
308969/2014-6; and CAPES, {\it Programa Est\'agio S\^enior no Exterior},
process number 88881.120856/2016-01.

This research has  made use of the SAO/NASA  Astrophysics Data System,
and the NASA/IPAC Extragalactic  Database (NED).  Funding for the SDSS
and  SDSS-II was  provided  by  the Alfred  P.  Sloan Foundation,  the
Participating  Institutions,  the  National  Science  Foundation,  the
U.S.  Department  of  Energy,   the  National  Aeronautics  and  Space
Administration, the  Japanese Monbukagakusho, the  Max Planck Society,
and  the Higher  Education  Funding  Council for  England.  A list  of
participating  institutions can  be obtained  from the  SDSS  Web Site
http://www.sdss.org/.











\bsp	
\label{lastpage}
\end{document}